\documentclass{svmult}

\usepackage{makeidx}         % allows index generation
\usepackage{graphicx}        % standard LaTeX graphics tool
\usepackage{multicol}        % used for the two-column index
\usepackage[bottom]{footmisc}% places footnotes at page bottom
\makeindex             % used for the subject index
\newcommand{\mueof}{\mu_{_{\mbox{\tiny EOF}}}}
\newcommand{\veof}{v_{_{\mbox{\tiny EOF}}}}

\newcommand{\gDPD}{\gamma_{{\mbox{\tiny DPD}}}}

\begin{document}

\title*{Mesoscopic simulations of polyelectrolyte electrophoresis in nanochannels}
\author{Jens Smiatek\inst{1}%{\footnote{Present adress: Institute of Physical
    %Chemistry, University of M{\"u}nster, Corrensstrasse 28/30, 48149
    %M{\"u}nster}}
\and
Friederike Schmid\inst{2}}
\institute{Institut f{\"u}r Physikalische Chemie, Westf{\"a}lische
  Wilhelms-Universit{\"a}t M{\"u}nster, D-48149
M{\"u}nster 
\texttt{jens.smiatek@uni-muenster.de}\\
\and{Institut f{\"u}r Physik, Johannes-Gutenberg Universit{\"a}t-Mainz,
  D-55699 Mainz \texttt{friederike.schmid@uni-mainz.de}}}
% used in your e-mail or web addressMDT07
%
\maketitle
\abstract{
We present the results of mesoscopic dissipative particle dynamics (DPD) simulations 
of coupled electrohydrodynamic phenomena on the micro- and nanoscale.
The effects of electroosmotic flow and slippage combined with polyelectrolyte
electrophoresis are investigated in detail, taking full account of hydrodynamic and
electrostatic interactions. Our numerical results are in excellent agreement with analytical calculations.}

\section{Introduction}
\label{sec:1}
Microfluidic devices like bio-MEMS (micro-electronical-mechanical-systems) and
bio-NEMS (nano-electronical-mechanical-systems) have attracted broad interest
over the last years due to their huge potential in biotechnology. 
The flow profiles in such micro- or nanosized devices are strongly influenced 
by the properties of the boundaries due to the large surface-to-volume ratio 
in these systems.  Surface characteristics like the wetting behavior
and/or slippage have a dramatic effect on the microscopic flow, leading to 
sometimes unexpected behavior.\\ 
One particularly important mechanism is electroosmotic transport: in contact 
with a liquid, many materials commonly used in nanotechnology ({\em e.g.}, 
polydimethylsiloxane (PDMS)) become charged due to ionizations of surface 
groups \cite{Israelachvili}. As a consequence, surfaces are often covered by
a compensating counterion layer \cite{Hunter}. In an external electric field, the 
ions are driven in one direction, dragging the surrounding solvent with them.
As a result, a flow is induced in the fluid, the electroosmotic flow (EOF). 
This electrokinetic effect has numerous consequences. For example, it alters 
drastically the migration dynamics of mesoscopic objects like polyelectrolytes 
or colloids \cite{Viovy00}. In microchannels, the EOF generated at the channel walls 
induces a total net flow, which is technologically attractive because it
can be controlled and manipulated more easily on the submicrometer scale 
than pressure- or shear-driven flow. \\
One important application of microchannels is to separate different fragments of
biological matter like DNA by their length for sequencing or further manipulation.
High molecular weight polyelectrolytes cannot be separated by naive electrophoresis 
in free salt solution due to the fact that the electrophoretic mobility becomes
length independent for long chains \cite{Viovy00}. In many separation methods,
the polyelectrolytes are driven through micro- or nanostructured environments
to overcome this problem, {\em e.g.}, a disordered gel (in gel electrophoresis), 
or structured microchannels. The presence of boundaries alters the dynamical
behavior of the macromolecules drastically\cite{Mathe07}, and their dynamical behavior
results from a complex interplay of electrostatics, hydrodynamics, and
confinement effects on the molecules.  \\
Our research focuses on the investigation of the explicit coupling of 
electrohydrodynamic effects in the nanometer scale in confined geometries
by coarse-grained mesoscopic simulations \cite{Smiatek09,Smiatek10}.
We use Dissipative Particle Dynamics (DPD), which is a popular mesoscopic
method in mesoscopic simulations. The results for polyelectrolyte 
electrophoresis in small microchannels are related to the experiments
published by Math\'{e} {\it et al.~}in \cite{Mathe07}. In particular,
we investigate in detail the influence of the electroosmotic flow on the 
total polyelectrolyte mobility. We find an excellent agreement between theory 
and numerical results.

\section{Dissipative Particle Dynamics}
\label{sec:4}
\label{sec:DPD}
Dissipative Particle Dynamics (DPD) was originally developed by Hoogerbrugge and
Koelman \cite{Hoo92} as a combination of Lattice
Gas Automata characteristics and Molecular Dynamics methods. 
Compared to atomistic Molecular Dynamics simulations,
this method gives access to much longer time- and length scales and is
therefore suited to study the long-time behaviour of soft matter systems and
transport phenomena.
It is coarse-grained, momentum-conserving, and creates a well-defined canonical ensemble.\\ 
The basic
DPD equations are given by
the forces on one particle, which involve two-particle interactions that are given
by
\begin{equation}
\label{eq:dpd}
  {\vec{F}}_{i}^{DPD}=\sum_{i\not={j}}{\vec{F}}_{ij}^{C}+{\vec{F}}_{ij}^{D}+{\vec{F}}_{ij}^{R}
\end{equation}
with a conservative force ${\vec{F}}_{ij}^{C}$
\begin{equation}
  {\vec{F}}_{ij}^{C} = -{\vec\nabla}_{ij}U_{ij}(r_{ij}),
\end{equation}
a dissipative force ${\vec{F}}_{ij}^{D}$ which reads
\begin{equation}
\label{eq:DPD}
  {\vec{F}}_{ij}^{D} =
  -\gamma_{DPD}\omega_{D}(r_{ij})(\hat{r}_{ij}\cdot{\vec{v}}_{ij})\hat{r}_{ij}
\end{equation}
with the friction coefficient $\gamma_{DPD}$ 
and an additional random force $\vec{F}_{ij}^{R}$ yielding 
\begin{equation}
\label{eq:random}
  {\vec{F}}_{ij}^{R} = \sigma\omega_{R}(r_{ij})\check{\zeta}_{ij}\hat{r}_{ij}.
\end{equation}
where the weighting function is given by
\begin{equation}
\omega_{D}(r_{ij})=[\omega_{R}(r_{ij})]^{2}\equiv\omega_{DPD}(r_{ij}),
\end{equation}
such that a canonical ensemble is sampled at equilibrium\cite{Gro97,Esp95}.
In DPD simulations, the conservative forces are often taken to have a certain soft shape \cite{Gro97}.
Here we only use the DPD {\em thermostat} as described above.
The random number $\check{\zeta}$ has zero mean and unit variance and
is symmetric, $\check{\zeta}_{ij}=\check{\zeta}_{ji}$, to ensure the conservation 
of momentum and the weighting function is arbitrary and often chosen 
linear \cite{Gro97}. It depends on the interparticle distance $r_{ij}$ and
the cut-off radius $r_c$.
The strength of the interaction is steered by the distance of the particles with
 \begin{eqnarray}
      \omega_{DPD}(r_{ij})=\left \{
        \begin{array}{cc}
          1-\frac{r_{ij}}{r_c} & :  r_{ij} < r_c \\
          0 & :  r_{ij} \geq r_c 
        \end{array}
      \right.
    \end{eqnarray}
while the amplitude of the Gaussian white noise in Eqn.~(\ref{eq:random}) is given by 
\begin{equation}
\sigma^{2}=2\gamma_{DPD}{k}_{B}T
\end{equation}
with the Boltzmann constant $k_B$ and the temperature $T$.
Eqn.~(\ref{eq:dpd}) can be integrated by an ordinary Molecular Dynamics integration scheme
like the Velocity-Verlet algorithm \cite{Fre01} which is used in the {\sf
  ESPResSo} package.

\section{The software package {\sf ESPResSo}}
All simulations in this work have been carried out using extensions of the
software package {\sf ESPResSo} (An {\bf E}xtensible {\bf S}imulation {\bf P}ackage for
{\bf Res}earch on {\bf So}ft matter) \cite{Espresso1}.
{\sf ESPResSo} was mainly developed for coarse-grained mesoscopic
simulation approaches. One of the advantages of this program is its high performance
MPI-parallelization implemented for simulations on supercomputers. Users can
furthermore change
and extend the program code to adopt it for their own purposes.
{\sf ESPResSo}
incorporates several simulation techniques like
Lattice-Boltzmann, Dissipative Particle Dynamics, Stochastic Dynamics as well
as pure Molecular Dynamics techniques. Another feature are the several implemented 
electrostatic algorithms like MMM1D, MMM2D, MMM3D, P3M, ELC and screened
Debye-H{\"u}ckel potentials, which allow to choose between the fastest
calculation methods available.
In summary, {\sf ESPResSo} provides a well founded basis code for high performance
computing on parallel clusters. 
The steering of the simulations is based on a {\sf TCL} (Tool Command
Language) script. For starting a simulation, no explicit knowledge
of implementation details is needed.
Even newcomers in the
methods of computer simulations can succesfully run a simulation after a short
time.\\ 
Several
tools for analysis are additionally included in the program. 
{\sf ESPResSo} is under public license and free to download
\cite{Espresso1}. Users that develop new ideas are invited to submit their
source code, written in the programming language {\sf C} to be incorporated after
testing in the newest release version. The development of {\sf ESPResSo}
continues and a number of extensions is already planned.\\
We ran our simulations on the NEC SX-8 Cluster at the High Performance
Computing Center Stuttgart. The number of computing nodes has normally been
chosen to 4 which corresponds to 32 CPUs. Each job has got a typical runtime of 4 hours
by using the concept of setting checkpoints.

\section{Polyelectrolyte electrophoresis in microchannels}
\label{sec:2}
\subsection{General theory}
We consider for simplicity a planar slit channel with
identical walls at $z = \pm L/2$, exposed to an external electric field $E_x$
in the $x$ direction. The electrostatic potential $\Phi(x,y,z)$ then takes the
general form $\Phi(x,y,z) = \psi(z) + E_x \: x + \mbox{const.}$ where we can
set $\psi(0)=0$ for simplicity. The electrolyte in the channel is taken to
contain $n$ different ion species $i$ with local number density $\rho_i(z)$ and
valency $Z_i$, which results in a net charge density $\rho(z) = \sum_{i=1}^n
(Z_i e) \rho_i(z)$. The electric field then generates a force density $f_x(z) =
\rho(z) E_x$ in the fluid.  Comparing the Poisson equation for the
electrostatic potential $\psi$,
\begin{equation}
 \label{eq:poisson}
 \frac{\partial^2 \psi(z)}{\partial z^2} = - \frac{\rho(z)}{\epsilon_r}
\end{equation}
(where $\epsilon_r$ is the dielectric constant), with the Stokes equation
\begin{equation}
 \label{eq:stokes}
 \eta_s \frac{\partial^2 v_x(z)}{\partial z^2} = - f_x(z) = - \rho(z) E_x
 \end{equation}
(with the shear viscosity $\eta_s$), one finds immediately $\partial_{zz}
v_x(z) = \partial_{zz} \psi(z) \:(\epsilon_r \: E_x/\eta_s) $.  For symmetry
reasons, the profiles $v_x$ and $\psi$ must satisfy the boundary condition
$\partial_z v_x|_{z=0} = \partial_z \psi|_{z=0} = 0$  at the center of the
channel. This gives the relation
\begin{equation}
 \label{eq:vx}
 v_x(z) = \frac{\epsilon_r \: E_x}{\eta_s} \psi(z) + \veof,
\end{equation}
where we have used $\psi(0)=0$ and identified the fluid velocity at the center
of the channel with the EOF velocity, $v_x(0) = \veof$. We further define
$\psi_B := \psi(\pm z_B)$ (for no-slip boundaries, $\psi_B$ is the so-called
Zeta-Potential \cite{Hunter,Smiatek10}).\\ 
The roughness of the channel boundaries is included in the partial slip boundary condition.
\begin{equation}
\label{eq:partial_slip}
\delta_B \: \: \partial_{z} v(x)|_{{z}_B} =  v_{x}(z)|_{{z}_B},
\end{equation}
where $v_x(z)$ denotes the component of the velocity in x-direction evaluated at the position ${z}_B$ of the so-called ``hydrodynamic boundary''.
This boundary condition is characterized by two effective parameters, namely
(i) the slip length $\delta_B$ and (ii) the hydrodynamic
boundary ${z}_B$. We note that the latter does not necessarily coincide
with the physical boundary.
Inserting the partial-slip boundary condition
for the flow, Eqn.~(\ref{eq:partial_slip}), we finally obtain the following simple
expression for the electroosmotic mobility,
\begin{equation}
 \label{eq:mueof2}
 \mueof = {\veof}/{E_x}
 = \mueof^0 \: (1 + \kappa \: \delta_B),
\end{equation}
where we have defined the inverse 'surface screening length'
\begin{equation}
 \label{eq:kappa}
 \kappa := \mp \frac{\partial_z \psi}{\psi} \large|_{z=\pm z_B},
\end{equation}
and $\mueof^0$ is the well-known Smoluchowski result \cite{Hunter} for the
electroosmotic mobility at sticky walls,
\begin{equation}
 \label{eq:mueof0}
 \mueof^0 = - \epsilon_r \: \psi_B/ \eta_s.
\end{equation}

The remaining task is to determine the screening parameter $\kappa$. If the
surface charges are very small and the ions in the liquid are uncorrelated, it
can be calculated analytically within the linearized Debye-H\"uckel theory
\cite{Israelachvili}. The Debye-H\"uckel equation for the evolution of the potential
$\psi$ in an electrolyte solution reads $\partial_{zz} \psi = \kappa_D^2 \psi $
with the inverse Debye-H\"uckel screening length
\begin{equation}
 \label{eq:kappa_D}
 \kappa_D = \sqrt{\frac{\sum_{i=1}^n (Z_i e)^2 \rho_{i,0}}
 {\epsilon_r \: k_{_{B}}T}},
\end{equation}
where $\rho_{i,0}$ is the density of ions $i$ far from the surface. It is
solved by an exponentially decaying function,
\begin{equation}
 \psi(z) \propto (e^{\kappa_D z} + e^{- \kappa_D z} - 2).
\end{equation}
Inserting that in Eq.~(\ref{eq:kappa}), one finds $\kappa = \kappa_D$, {\em
i.e.}, the surface screening length is identical with the Debye screening
length. 

Unfortunately, the range of validity of the Debye-H\"uckel theory is limited,
it breaks down already for moderate surface potentials $\psi_B$ and/or for
highly concentrated ion solutions. Nevertheless, the exponential behavior often
persists even in systems where the Debye-H\"uckel approximation is not valid.
For high ion concentrations a Debye-H\"uckel-type approximation can still be used
in a wide parameter range, if $\kappa_D$ is replaced by a modified effective
screening length \cite{Boroudjerdi}.  For high surface
charges, analytical solutions are again available in the so-called 'strong
coupling limit', where the profiles are predicted to decay exponentially with
the Gouy-Chapman length \cite{Moreiraa}. This limit is very special and rarely
encountered. At intermediate coupling regimes, the decay length must be
obtained empirically, {\em e.g.}, by fitting the charge distribution $\rho(z)$
to an exponential behavior, which is characterized by the same exponential
behavior than $\psi(z)$ by virtue of the Poisson equation,
\begin{equation}
  \sum_{i=q}^n (Z_i e) \: \rho_i(z)
  \propto  \frac{\partial^2 \psi(z)}{\partial z^2} \propto
  (e^{\kappa z} + e^{- \kappa z}).
\end{equation}
Assuming that the electrophoretic velocity of the polyelectrolyte
\begin{equation}
v_p(x) = \mu_e E_x
\end{equation}
is influenced by the electroosmotic mobility of the counterions $\mu_{EOF}$, a total mobility has to be defined
$\mu_t = \mueof +\mu_e$
which describes the overall mobility of the electrophoretic object.
Putting everything together, the total net electrophoretic mobility $\mu_t$
of a polyelectrolyte in the channel can be expressed in
terms of the electroosmotic mobility $\mueof$ as
\begin{equation}
\label{eq:eof_mobp}
  \frac{\mu_t}{\mueof}=1+\frac{\mu_e}{\mueof^0(1+ \kappa \: \delta_B)},
\end{equation}
where the ratio $\mu_e/\mueof^0$ depends only weakly on the ionic strength of
the electrolyte and the slip length of the surface.  The main effect of slippage
is incorporated in the factor $(1 + \kappa \: \delta_B)^{-1}$i \cite{Smiatek10}.

\subsection{Simulation details}
We have studied the electrophoresis of a single charged polymer of length $N=20$ in
electrolyte solutions, confined by a planar slit channel with charged walls.
All particles, polymer, solvent and ions, are modeled explicitly. We use a
simulation box of size ($12\sigma\times 12\sigma\times 10\sigma$) which is
periodic in $x$- and $y$-direction and confined by impermeable walls in the
$z$-direction. The walls repel the particles {\em via} a soft repulsive WCA
potential of range $\sigma$ and amplitude $\epsilon$. (Hence the
accessible channel width for the particles is actually $L_z = 8 \sigma$).  Ions
and monomers repel each other with the same WCA potential. In addition, chain
monomers are connected by harmonic springs
\begin{equation}
  U_{harmonic}=\frac{1}{2}{k}(r_{ij}-r_0)^2
\end{equation}
with the spring constant $k=25 \epsilon/\sigma^2$ and $r_0=1.0\sigma$.  Neutral
solvent particles have no conservative interactions except with the walls.
The wall contains immobilized, negatively charged particles at random
positions.  Every second monomer on the polyelectrolyte with 20 beads carries a negative
charge resulting in 10 charged monomers. The solvent contains the positive counterions
for the walls and the polyelectrolyte, and additional (positive and negative) salt ions.
All charges are monovalent, and the system as a whole is electroneutral. In addition to
their other interactions, charged particles interact {\em via} a Coulomb
potential with the Bjerrum length $\lambda_B=e^2/4\pi\epsilon_r
k_BT=1.0\sigma$, and they are exposed to an external field $E_x=-1.0
\epsilon/e\sigma$. Specifically, we have studied systems with a surface charge
density of $\sigma_A=-0.208e\sigma^{-2}$ which corresponds to $30$ charged particles per wall.
In a recent publication \cite{Smiatek09}, we have shown that this corresponds
to the 'weak-coupling regime', {\em i.e.}, the regime where the Poisson-Boltzmann theory
is valid.  The electrostatic coupling constant \cite{Moreiraa}
$\Xi = 2 \pi Z^3 \lambda_B^3 \sigma_A $ ($Z=1$ is the valency of the cations),
which gives the strength of electrostatic interactions between the surface and the
ions compared to thermal energy, is close to unity, $\Xi \sim 1.3$. In this earlier
work, we have also studied the effect of using homogeneously charged wall instead
of discrete embedded charge, and found the differences to be negligible \cite{Smiatek09}.
The use of discrete embedded charges has practical advantages in our simulation code, which
is why we use them here. The total counterion density was
$\rho = 0.06\sigma^{-3}$ and the salt density varied between
$\rho_s=$0.05625, 0.0375, 0.03, 0.025, and $0.015 \sigma^{-3}$. In molar units, this
corresponds to 0.272, 0.181, 0.145, 0.121 and 0.072 mol/l, if we identify
$\lambda_B \approx 0.7$ nm, {\em i.e.}, the Bjerrum length in water at room temperature
\cite{Viovy00}.

We use DPD simulations with a friction coefficient
$\gDPD=5.0\sigma^{-1}(m\epsilon)^{1/2}$.  (More precisely, we only use the
DPD {\em thermostat}, not the soft conservative DPD forces. All conservative
forces in our system are either WCA forces, spring forces, or Coulomb forces
as explained above). The density of the solvent particles
was $\rho=3.75\sigma^{-3}$, and the temperature of the system was
$T=1.0\epsilon/k_B$. For these parameters, the shear viscosity of the DPD fluid
-- as determined by fitting the amplitude of Plane Poiseuille flows
\cite{Smiatek} -- is given by
$\eta_s=(1.334\pm0.003)\sigma^{-2}(m\epsilon)^{1/2}$.  The DPD timestep was
$\delta t=0.01\sigma(m/\epsilon)^{1/2}$.

Tunable-slip boundary conditions were used with friction coefficients
$\gamma_L=$0.1, 0.25, 0.5, 0.75, 1.0, and $6.1 \sigma^{-1}(m\epsilon)^{1/2}$.
The range of the viscous layer was $z_c=2.0\sigma$.  Only the solvent particles
interact with the tunable-slip boundaries.  By performing Plane Poiseuille and
Plane Couette flow simulations with the above given parameters, the slip length
$\delta_B$ and the hydrodynamic boundary positions $z_B$ can be determined
independently \cite{Smiatek}.  The hydrodynamic boundary position is found at
$|z_B|=(3.866\pm 0.266)\sigma$ in all simulations. 

\subsection{Results}
\label{sec:nr}
\subsubsection{Polyelectrolytes in absence of external fields}
To understand the behaviour of polyelectrolytes in salty solution we first studied the dynamics in absence of external fields.
The general theory \cite{Viovy00} implies that hydrodynamic interactions between the monomers are important to describe the dynamic behaviour adequately.
This results in the so called Zimm dynamics \cite{Viovy00}. If external electric fields are present, hydrodynamic interactions between the charged monomers
are screened due to the presence of salt. The two rivaling effects which are the electrophoresis of the polyelectrolyte as well as the electroosmosis of the
mobile counterions in opposite direction lead to a crucial screening of hydrodynamic interactions which is described by Rouse dynamics \cite{Viovy00}.\\
A powerful tool to investigate the underlying dynamics in experiments as well as in computer simulations is given by the dynamic version of the structure factor 
which is defined by
\begin{equation}
S(\vec{k},t)=\frac{1}{N}\sum_{i,j}<e^{i\vec{k}(\vec{R}_i(t)-\vec{R}_j(t_0))}>
\end{equation}
with the actual monomer position $\vec{R}_i$ respectively $\vec{R}_j$.
For the inverse length scale $1/R_g\ll k \ll 1/a_0$ with the gyration radius $R_g$ \cite{Doi86} and the smallest microscopic length $a_0$ and the finite time interval
$t_b\ll t \ll\tau$ after the ballistic time $t_b$ and the longest relaxation time $\tau$, the dynamic
structure factor obeys the following scaling relation \cite{DuenKrem93}
\begin{equation}
  \label{eq:scale_sk_t}
  S(k,t)=S(k,0)f(k^zt)
\end{equation}
which depends on the the parameter $z$. 
This parameter differentiates between the different regimes. 
For the Zimm dynamics it is given by
\begin{equation}
  \label{eq:zZ}
  z=3\quad \mbox{(Zimm-Regime)}
\end{equation}
and for Rouse dynamics
\begin{equation}
  \label{eq:zR}
  z=2+1/\nu \quad\mbox{(Rouse-Regime)}
\end{equation}
with the Flory paramter $\nu$ which describes the inverse fractal dimension of the polyelectrolyte \cite{Doi86}.\\
We have compared an uncharged polymer and a half charged polyelectrolyte in a salt solution of concentration $\rho_s=0.05\sigma^{-3}$ and a solvent density of 
$\rho=3.0\sigma^{-3}$ with free periodic boundary conditions in all dimensions. All the other parameters are in agreement to the above described simulation details.
\begin{figure}
  \includegraphics[width=\textwidth]{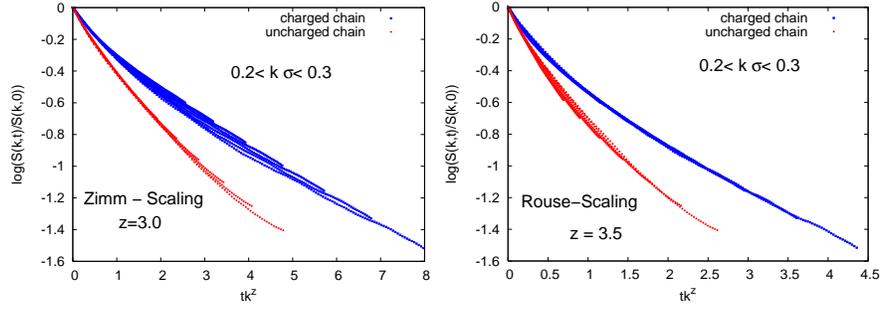}
\caption{ 
  Dynamic structure factor S(k,t) for an uncharged chain (red) in
      correspondence to a half charged chain (blue) for $0.2<k\sigma<0.3$ and
      salt concentration $\rho_s=0.05\sigma^{-3}$. The
      timescale for the uncharged chain is $0<t/\tau<180$ in contrast to $0<t/\tau<300$ for
      the half charged polyelectrolyte. Both chains consist of $N=50$ monomers.
      Left side: Zimm-scaling with $z=3$.
      Right side: Rouse-scaling with $z=3.5$.
}
\label{fig:6}
\end{figure}
Fig.~\ref{fig:6} presents our simulation results for an uncharged chain and the polyelectrolyte. Our results indicate that although external fields are absent, 
polyectrolytes can be described best by Rouse dynamics whereas uncharged polymers can be described by Zimm dynamics. More results on that topic can be found in
\cite{phd,manuscript}.  

\subsubsection{Polyelectrolyte electrophoresis in microchannels}
\begin{figure}
  \includegraphics[width=\textwidth]{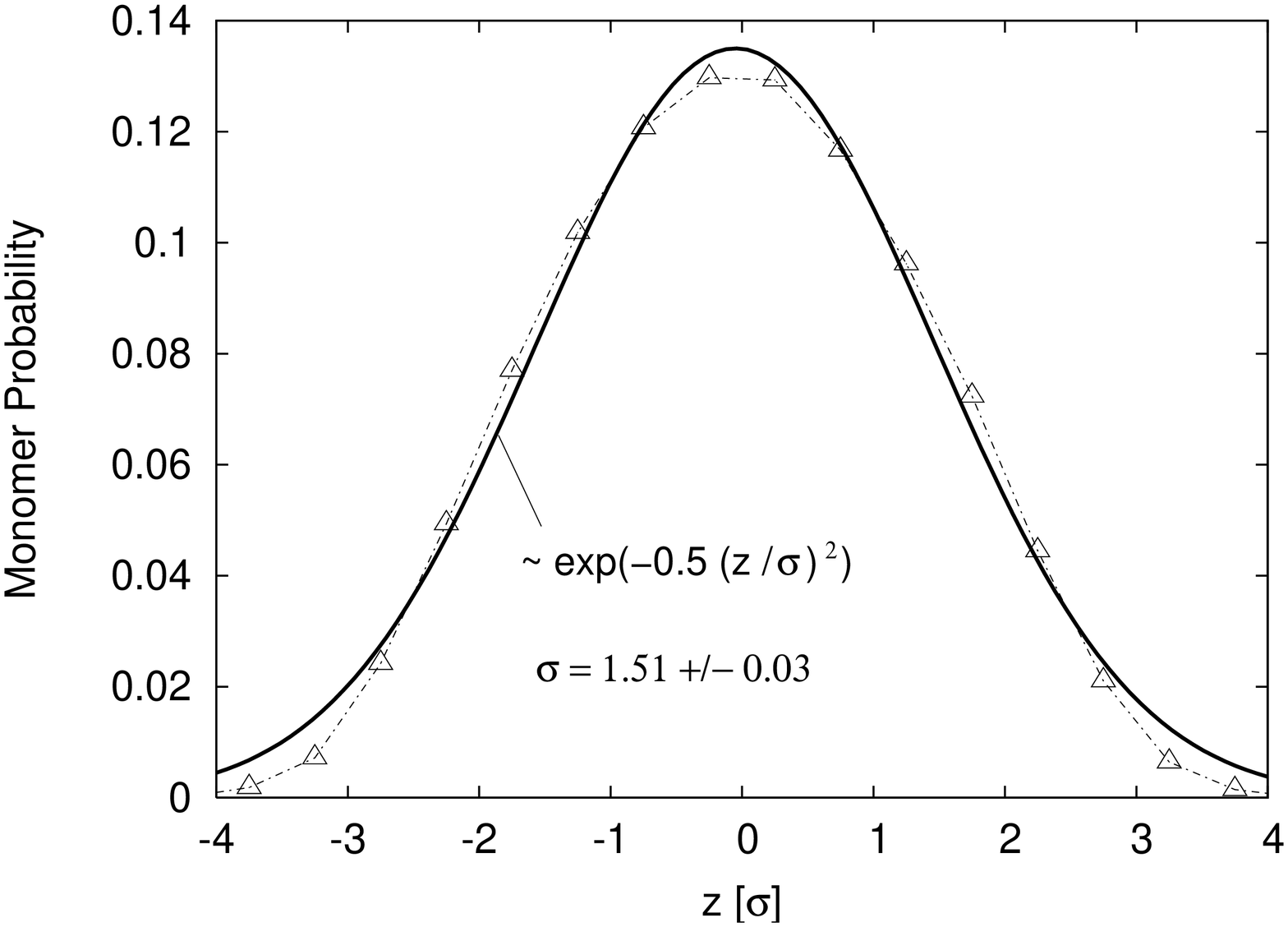}
\caption{Monomer appearance probability inside the channel for a salt
  concentration of $\rho_s=0.05625\sigma^{-3}$.
}
\label{fig:mono20dens}
\end{figure}

In this section our numerical results are presented which indicate the drastic
influence of the electroosmotic flow on the total mobility of the
polyelectrolyte. Larger slip lengths even enhance this effect.\\
\begin{figure}
  \includegraphics[width=\textwidth]{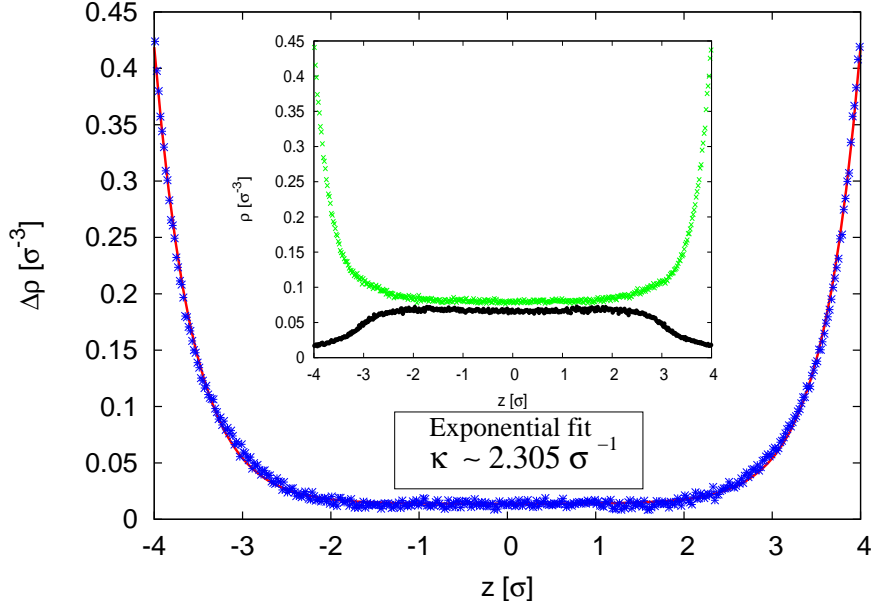}
\caption{Distribution of the ionic difference $\Delta
  \rho=\rho_c-\rho_a$ for an exemplary salt concentration of $\rho_s=
  0.05625\sigma^{-3}$ with a surface ion density of
  $\sigma_s=0.208\sigma^{-2}$. The red line corresponds to the exponential
  fit of Eqn.~(\ref{eq:fit}) with an effective inverse screening length of $\kappa=2.305\pm0.025\sigma^{-1}$.
  {\bf Inset:} Distribution of cations (salt cations and counterions) with green symbols and anions
  (salt anions) with black symbols for an the exemplary salt concentration for the
  above given parameters.}
\label{fig:ionscomb}
\end{figure}
Fig.~\ref{fig:ionscomb} presents the results of the ionic distribution for a
salt concentration of $\rho_s=0.05625\sigma^{-3}$ and a counterion density of $\rho=0.0525\sigma^{-3}$
with a surface ion density of $\sigma_s=0.208\sigma^{-2}$ in presence of the polyelectrolyte. Due to the large
number of cations in comparison to the anions, the cation density in the middle of the channel
is increased (inset of Fig.~(\ref{fig:ionscomb})).
Calculating the ionic difference $\Delta\rho=\rho_c-\rho_a$ as the difference of
the cationic and anionic density yields
\begin{equation}
\frac{\partial^2}{\partial z^2}\psi(z)=
  -\sum_i^N\frac{Z_ie}{\epsilon}\rho_i(z)=-\frac{Ze}{\epsilon}\Delta \rho
\end{equation}
with the unique valency $|Z|$ \cite{Smiatek10}. As we have mentioned in section \ref{sec:2}, the
application of the Debye-H{\"u}ckel theory for non electroneutral bulk systems
is not valid and has to be replaced by an effective inverse screening length $\kappa$.
Thus we have used the following fit function
\begin{equation}
\label{eq:fit}
  \Delta \rho = \Delta\rho_0(e^{-\kappa z}+e^{\kappa z})+c
\end{equation}
to determine the effective inverse screening length $\kappa$,
with the fit parameters $\Delta\rho_0,c$ and $\kappa_D$. The red
line in Fig.~\ref{fig:ionscomb} shows that the fit describes the ionic difference
adequately.  The fitted parameter values of $\kappa$ for the different salt
concentrations are presented in Tab.\ref{tab:2}.
Significant differences between the inverse screening lengths
are only observed for
salt concentrations $\rho_s\geq 0.0375\sigma^{-3}$. Thus, for lower
salt concentrations, the screening of electrostatic interactions is mainly
effectuated by the much larger number of counterions instead of the salt ions \cite{Smiatek10}.\\
\begin{table}[t]
\caption{Fitted inverse screening lengths $\kappa$
  for the different salt concentrations $\rho_s$ and a fixed counterion
  density of $\rho=0.0525\sigma^{-3}$.}
\label{tab:2}       % Give a unique label
\begin{center}
\begin{tabular}{cc}
\hline\noalign{\smallskip}
$\rho_s[\sigma^{-3}]$ & $\kappa [\sigma^{-1}]$\\
\noalign{\smallskip}\hline\noalign{\smallskip}
0.015 & 1.996 $\pm$ 0.041 \\
\noalign{\smallskip}\hline\noalign{\smallskip}
0.0225 & 2.011 $\pm$ 0.049 \\
\noalign{\smallskip}\hline\noalign{\smallskip}
0.03 & 1.983 $\pm $0.041 \\
\noalign{\smallskip}\hline\noalign{\smallskip}
0.0375 & 2.182 $\pm $ 0.047 \\
\noalign{\smallskip}\hline\noalign{\smallskip}
0.05625 & 2.305 $\pm $ 0.025 \\
\noalign{\smallskip}\hline
\end{tabular}
\end{center}
\end{table}
\begin{figure}
  \includegraphics[width=\textwidth]{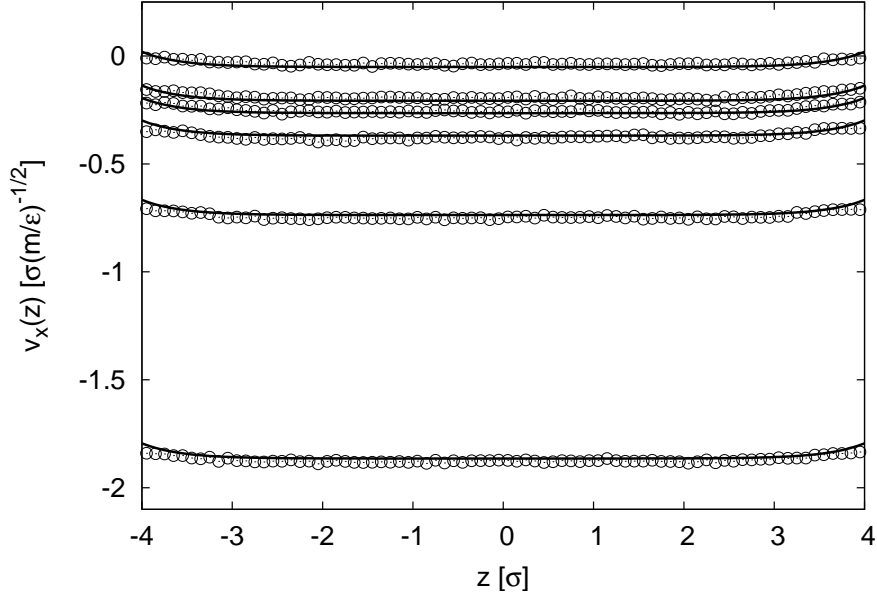}
  \caption{Exemplary flow profiles for a salt concentration
    $\rho_s=0.05625\sigma^{-3}$ for varying slip lengths (from bottom to top:
    $\delta_B=(14.98, 5.66, 2.63, 1.77, 1.29, 0.00)\sigma$.) The black lines are
    the integrated theoretical predictions in terms of the Stokes equation
    (\ref{eq:stokes}) with a fitted inverse screening
    length of $\kappa=2.305 \sigma^{-1}$. }
\label{fig:solvflow}
\end{figure}
The corresponding electroosmotic flow profiles for an exemplary salt concentration of
$\rho_s=0.05625\sigma^{-3}$ are shown in Fig.~\ref{fig:solvflow}. The
different magnitudes correspond to varying slip lengths. 
Larger slip lengths enhance the flow profile
drastically. All points are in good agreement to the integrated
analytical expression of Eqn.~(\ref{eq:fit}) in terms of the Stokes equation
(\ref{eq:stokes}) with partial-slip boundary
conditions. Thus the description of the electroosmotic flow in terms of the
Stokes theory is valid. \\
\begin{figure}
  \includegraphics[width=\textwidth]{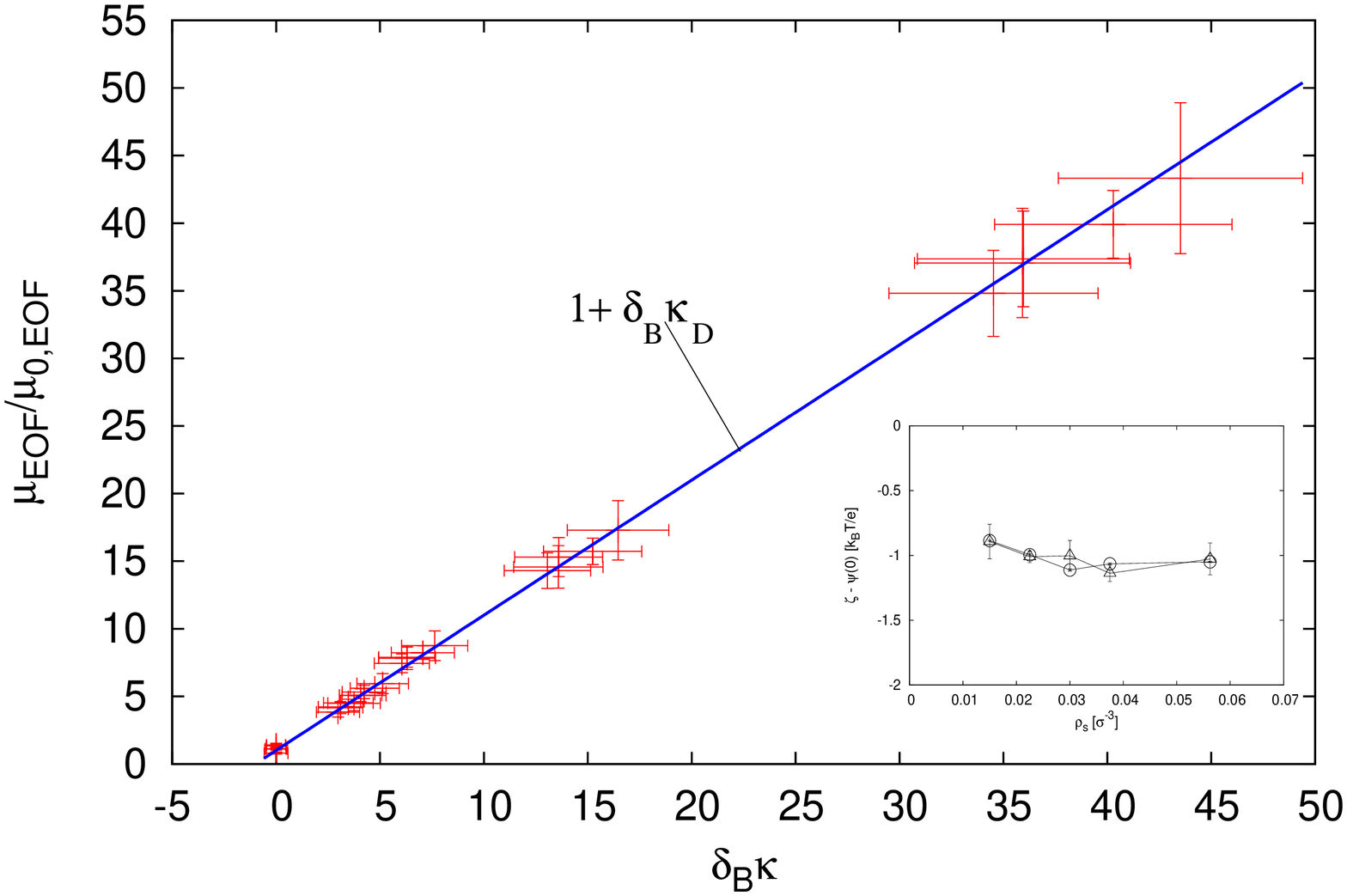}
\caption{Ratio $\mu_{_{EOF}}/ \mu_{_{0,EOF}}$ plotted against $\delta_B\kappa$
  for the different salt concentrations and screening lenghts given in
  Tab.~\ref{tab:2} with counterions of density $\rho=0.0525\sigma^{-3}$.
  The
  blue line is the theoretical prediction of Eqn.~(\ref{eq:mueof2}) with slope
  $1+\delta_B\kappa$.}
\label{fig:solvcscale}
\end{figure}
Fig.~\ref{fig:solvcscale} compiles our numerical results for the electroosmotic
mobility for all salt concentrations and slip lengths. They are in very
good agreement to the theoretical prediction of Eqn.~(\ref{eq:mueof2}), where
$\mu_{_{0,EOF}}$ has been determined independently by a linear regression for
each salt concentration. It is worth noting that
the presence of the polyelectrolyte does not perturb the amplitude
of the electroosmotic flow. \\
After investigating the electroosmotic flow of the solvent, we
focus on the dynamics of the polyelectrolyte.
The monomer distribution for a salt concentration of
$\rho_s=0.05625\sigma^{-3}$ is presented in
Fig.~\ref{fig:mono20dens}. The fit shows that the distribution is dominated
by a peak in the middle of the channel with a
variance of $Var \sim 2.28\sigma$.
Thus the main drag of the electroosmotic flow on the polyelectrolyte is exerted 
in the middle of the channel, and the rapid variations of the
electroosmotic flow in close vicinity to the boundaries (Fig.~\ref{fig:solvflow}) 
have little influence on the polyelectrolyte mobility. 
Estimating the total mobility by assuming a constant plug-like
flow profile for the fluid in the middle of the channel from $z=-1.52\sigma$
to $z=1.52\sigma$ due to the variance of the monomer distribution
therefore seems reasonable.\\
The influence of the electroosmotic flow on the polyelectrolyte can be
investigated by regarding static properties like the radius
of gyration $R_g^2=(1/2N^2)\sum_{i,j=1}^N<(\vec{R}_i-\vec{R}_j)^2>$
and the end-to-end radius with $R_e^2 = <(\vec{R}_N-\vec{R}_1)^2>$ \cite{Doi86}.
\begin{table}
\caption{Radius of gyration $R_g$ and end to end radius $R_e$ for a
  polyelectrolyte with $N=20$ monomers for different salt concentrations $\rho_s$.}
\label{tab:3}       % Give a unique label
\begin{tabular}{lll}
\hline\noalign{\smallskip}
$\rho_s[\sigma^{-3}]$ & $R_g [\sigma]$ & $R_e[\sigma]$ \\
\noalign{\smallskip}\hline\noalign{\smallskip}
0.015 & $3.2218\pm 0.047$  & $10.6480\pm0.0314$ \\
\noalign{\smallskip}\hline\noalign{\smallskip}
0.0225 & $3.1661\pm0.0041$ & $10.2736\pm0.0266$ \\
\noalign{\smallskip}\hline\noalign{\smallskip}
0.03 & $3.1486\pm0.0451$ & $10.1777\pm0.0292$ \\
\noalign{\smallskip}\hline\noalign{\smallskip}
0.0375 & $3.1279\pm0.0045$ & $10.0819\pm0.0287$ \\
\noalign{\smallskip}\hline\noalign{\smallskip}
0.05625 & $3.0825\pm0.0045$ & $9.8331\pm0.0280$ \\
\noalign{\smallskip}\hline
\end{tabular}
\end{table}
The values of these parameters are shown in Tab.~\ref{tab:3}. Both
properties decrease with larger salt concentration due to a more pronounced
screening of electrostatic interactions which is in accordance to standard
theories \cite{Viovy00}. Although the values for the end to end radius
are quite large, in all cases the box dimensions in the unconfined
directions with $b_{x,y}=12\sigma$ are larger than the maximal average extension of
the chain. Possible explanations for these large ratios between the
end-to-end and gyration radii could be shear-induced elongation
\cite{Boroudjerdi} or direct squeezing due to the presence of the channel
walls.\\
\begin{figure}
  \includegraphics[width=\textwidth]{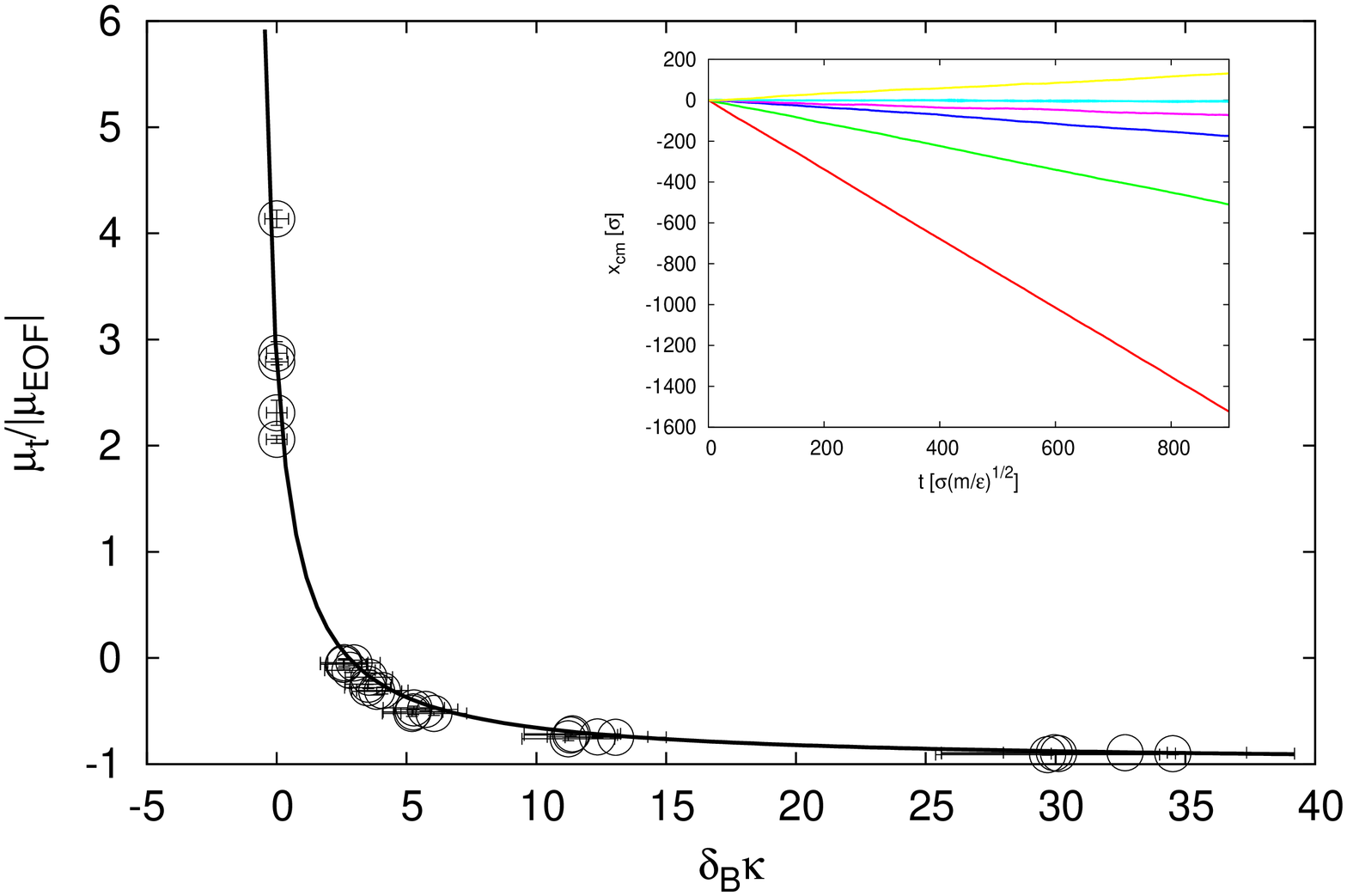}
\caption{Ratio $\mu_t/|\mu_{_{EOF}}|$ plotted against $\delta_B\kappa$ for all
  salt concentrations. The
  black line is the theoretical prediction of Eqn.~(\ref{eq:eof_mobp}) with
  absolut values of $|\mu_{_{EOF}}|$. In the
  limit $\delta_B\kappa\rightarrow \infty$, the total mobility of the
  polyelectrolyte is equal to the electroosmotic flow. The ratio
 $\mu_p/\mu_{_{EOF}}^0$ has been fitted to $-3.778\pm 0.128$. Negative values
  of $\mu_t/|\mu_{_{EOF}}|$ indicate absolute negative total mobilities of the
  polyelectrolyte.
  {\bf Inset:}
  Total displacement of the polyelectrolytes center of mass for different
  boundary conditions and a salt concentration of
  $\rho_s=0.05625\sigma^{-3}$. The total mobility becomes negative if the
  relation $|\mu_e|\ll|\mu_{_{EOF}}|$ is fulfilled. The lines correspond from 
  top to bottom
  to the slip lengths $\delta_B\approx (0.00, 1.292, 1.765, 2.626, 5.664,
  14.98)\sigma$. Thus larger slip lengths indirectly enhance the total mobilty of
  the polyelectrolyte.
}
\label{fig:mobscalefree}
\end{figure}
The total mobility of the polyelectrolyte for varying boundary conditions is
finally presented in Fig.~\ref{fig:mobscalefree}. The theoretical prediction
of Eqn.~(\ref{eq:eof_mobp}) is in good agreement with the numerical results with
the fitted ratio $\mu_e/\mu_{_{EOF}}^0=-3.778\pm 0.128$. With no-slip boundaries 
($\delta_B\approx 0$) one obtains ordinary behaviour where the polyelectrolyte
migrates in opposite direction to the electroosmotic flow. In the presence
of slip, the absolute mobility may become negative if the amplitude of the
electroosmotic flow exceeds a critical value given by a combination of the
inverse screening and slip lengths, and if the sign of the wall charges
is identical to the net charge of the polyelectrolyte. If the wall is
oppositely charged to the polyelectrolyte, slippage effects should even enhance the
total velocity of the polyelectrolyte.\\
The right inset of
Fig.~\ref{fig:mobscalefree} shows the total displacement of the chains center of mass
for various slip lengths and a salt concentration of
$\rho_s=0.05625\sigma^{-3}$. In nearly all cases
except for $\delta_B\approx 0$, the total
mobility of the polyelectrolyte is negative with
$|\mu_p|\ll|\mu_{_{EOF}}|$, indicating negative values of
$\mu_t/\mu_{_{EOF}}^0$. This can only be explained in
terms of the drag force of the electroosmotic flow and the surplus of cations
in close vicinity to the boundaries (Fig.~\ref{fig:ionscomb}) and is in
agreement to the results derived above.\\
To summarize, the assumptions and calculations of section \ref{sec:2} are in good
agreement to the presented numerical results in this section. The total mobility of the
polyelectrolyte can therefore be adequately described by Eqn.~(\ref{eq:eof_mobp}).
\section{Summary}
We have presented mesoscopic DPD simulations of polyelectrolyte
electrophoresis in narrow microchannels, taking full account of hydrodynamic
and electrostatic interactions.
We have shown that the product of the inverse
screening length $\kappa$ and the slip length $\delta_B$ massively
influences the electroosmotic flow and therefore the total mobility
of the polyelectrolyte. Thus the characteristics of the boundaries
have to be taken into account for a proper description of the polyelectrolyte
migration dynamics. For certain parameter sets, even a negative mobility
can be achieved.  All our numerical results are
in good agreement to the analytical derived results.\\
In summary, only a combination of electroosmotic, electrophoretic,
electrostatic and slippage effects does describe the total mobility of
polyelectrolytes in microchannels adequately. Our simulations indicate and
explain total negative mobilities due to boundary effects which also have been
observed in recent experiments \cite{Mathe07}.\\
The characteristics of the channel walls could be used to significantly enhance flow
profiles which offers the possibility to reduce
the time needed for polymer migration or separation techniques. This could be an important
aspect for future applications in microchannels or micropumps to accelarate
the measuring time in experiments.\\

\section{Acknowledgements}
We thank Christian Holm, Burkhard D{\"u}nweg, Ulf D. Schiller, Marcello Sega, Michael P. Allen
and Kai Grass for nice and fruitful discussions.
Furthermore we thank the HLRS in Stuttgart for computing time and the
Volkswagen Stiftung for funding.
%%%%%%%%%%%%%%%%%%%%%%%% referenc.tex %%%%%%%%%%%%%%%%%%%%%%%%%%%%%%
% sample references
%
%
% Use this file as a template for your own input.
%
%%%%%%%%%%%%%%%%%%%%%%%% Springer-Verlag %%%%%%%%%%%%%%%%%%%%%%%%%%

%
% BibTeX users please use
% \bibliographystyle{}
% \bibliography{}

%%% Local Variables: 
%%% mode: latex
%%% TeX-master: "referenc"
%%% End: 

%%%%%%%%%%%%%%%%%%%%%%%%%%%%%%%%%%%%%%%%%%%%%%%%%%%%%%%%%%%%%%%%%%%%%%

%%%%%%%%%%%%%%%%%%%%%%%%%%%%%%%%%%%%%%%%%%%%%%%%%%%%%%%%%%%%%%%%%%%%%%

\printindex
\end{document}